\documentclass[a4paper,12pt]{article}
\usepackage[english]{babel}
\usepackage[numbers,comma,square,compress]{natbib}
\usepackage{amssymb,amsmath,amsfonts,bm,enumerate}
\usepackage{graphicx}

\usepackage{color}
\definecolor{hlcolor}{rgb}{0.5,0,0.8}

\usepackage{ifpdf}
\ifpdf  
 \usepackage[pdftex,a4paper,hmargin=25mm,vmargin=25mm]{geometry}
 \usepackage[pdftex,colorlinks=true,allcolors=blue]{hyperref}
\else  
 \usepackage[dvips,a4paper,hmargin=25mm,vmargin=25mm]{geometry}
 \usepackage[dvips,colorlinks=true,allcolors=blue]{hyperref}
 \fi

\DeclareMathAlphabet{\mathcal}{OMS}{cmsy}{b}{n}

\newcommand{\email}[1]{\footnote{E-mail: \href{mailto:#1}{#1}}}

\def \sumint{\sum \!\!\!\!\!\!\!\! \int }
\begin{document}

\title{\bf\Large{Thermal Effects of Very Special Relativity QED}}
\author{\bf{R. Bufalo }$^{1}$\email{rodrigo.bufalo@ufla.br}~ \textbf{and} \bf{M. Ghasemkhani} $^{2}$\email{ghasemkhani@ipm.ir } \\
\textit{$^{1}$ \small Departamento de F\'isica, Universidade Federal de Lavras,}\\
\textit{ \small Caixa Postal 3037, 37200-000 Lavras, MG, Brazil}\\
\textit{\small $^{2}$  Department of Physics, Shahid Beheshti University,  G.C., Evin, Tehran 19839, Iran}\\
}

\maketitle
\date{}

\begin{abstract}
In this paper we compute the high-temperature effective Lagrangian for the quantum electrodynamics defined in the framework of very special relativity (VSR).
The main aspect of the VSR setting is that it modifies the gauge invariance, admitting now different types of interactions appearing in a nonlocal form.
In order to explore the richness of these new couplings, we employ the usual Matsubara imaginary-time formalism to evaluate the effective Lagrangian at one and two-loop order.
We illustrate the leading VSR Lorentz violation modifications by computing high-temperature internal energy density and establishing a comparison with the expected contribution.
\end{abstract}

\section{Introduction}
Despite the fact that many predictions of the Standard Model (SM) of particles are in accordance with experimental data obtained at particle colliders, there are many other physical phenomena that are not adequately explained by SM, consisting of the so-called physics beyond the Standard Model, e.g., neutrino masses, matter-antimatter asymmetry, quantum gravity, etc. \cite{Lykken:2010mc}.
Minimal modifications of SM or alternative models have been proposed in order to attempt to understand the fundamental origin of
one or more of the above problems.
The most appealing approach for the description of such phenomena involves
the addition of new degrees of freedom (d.o.f.), that are usually
incorporated by enforcing a symmetry principle.

In recent years, the number of precision tests scrutinizing the validity of exact symmetries in field theories have been increased significantly.
Since Lorentz invariance is a key symmetry in our description of nature, Lorentz violating models have reached an important landmark due to many systematic developments that were subject to precision tests
\cite{ref53,Jacobson:2005bg,Bluhm:2005uj,Ellis:2005wr}.
These studies are also motivated by our expectation that any deviation from these symmetries would signal manifestations of Planck-scale physics, once it is believed that they would not hold exactly in this energy regime  \cite{AmelinoCamelia:2008qg}.

Among the many Lorentz violating proposals applied to field theories, certainly those preserving the basic elements of special relativity are the most relevant ones, because although all classical tests of special relativity are still valid, these modified models allow the description of additional phenomena.
Within this context, a framework satisfying the above criteria is the  Cohen and Glashow very special relativity (VSR) \cite{Cohen:2006ky,Cohen:2006ir}.
The main aspect in the VSR proposal is that the laws of physics are not invariant under the whole Poincar\'{e} group but rather under subgroups of the Poincar\'{e} group preserving the basic elements of special relativity.
This approach is mainly motivated by the fact that a modified gauge symmetry is present in the VSR framework, admitting a variety of new gauge invariant interactions.

In the VSR framework, there are two subgroups fulfilling the aforementioned requirements, namely, the HOM(2) (with three parameters) and the SIM(2) (with four parameters). The former is the so-called Homothety group, generated by $T_1 = K_x + J_y$, $T_2 = K_y-J_x$, and $K_z$ ($\vec{J}$ and $\vec{K}$ are the generators of rotations and boosts, respectively).
The latter, called the similitude group SIM(2), is the HOM(2) group added by the $J_z$ generator.
Moreover, the symmetry groups SIM(2) and HOM(2) preserve the direction of a lightlike four-vector $n_{\mu}$ by a
simple scaling, in which it transforms as $n \to e^{\varphi} n$ under boosts in the z-direction.
This shows that theories defined under invariance of these subgroups present a preferred direction in Minkowski spacetime, where Lorentz violating terms are constructed as ratios of contractions of $n_{\mu}$ with other kinematic vectors \cite{Cohen:2006ky}.

As an illustration of the above discussion, one can write down a VSR-covariant Dirac equation in the form
\begin{equation}
\left(i\gamma^\mu \tilde{\partial}_\mu -m_e \right)\psi \left(x\right)=0,
\end{equation}
where the wiggle derivative operator is defined by $\tilde{\partial}_{\mu}=\partial_{\mu}+\frac{1}{2}\frac{m^{2}}{n.\partial}n_{\mu}$, carrying Lorentz violation in a nonlocal form, where $m$ sets the scale for the VSR effects and the preferred null direction is chosen as $n_{\mu}=\left(1,0,0,1\right)$.
Then, squaring the VSR-covariant Dirac equation we find
\begin{equation}
\left(\partial^\mu \partial_\mu +\mu^2 \right)\psi \left(x\right)=0,
\end{equation}
where $\mu^2=m_e^2+m^2 $.
This shows a basic result of VSR that conservation laws and the usual relativistic dispersion relation are preserved.
Actually,the main motivation for the VSR proposal was to provide a suitable framework where the neutrinos are massive that neither violate lepton number nor require additional sterile states \cite{Cohen:2006ir}.
Moreover, this idea can be extended to the photon, where the Abelian gauge field now transforms as $\delta A_\mu = \tilde{\partial}_\mu \Lambda$, the extended VSR gauge symmetry can be used in such a way to describe massive modes without changing the number of physical polarization states for the photon \cite{Cheon:2009zx}.
It is worth noting that massive modes of photons and their stability are a recurrent
subject of analysis in recent literature \cite{Heeck:2013cfa,Bonetti:2017pym}.

Many interesting theoretical and phenomenological aspects of VSR effects have
been extensively discussed \cite{Alfaro:2013uva,Bufalo:2015gja,Bufalo:2016lfq,Bufalo:2017yxs,Alfaro:2015fha,Nayak:2016zed,Dunn:2006xk,Alfaro:2019koq}.
However, no attention has been paid to the analysis of observable quantities of VSR field theories in thermodynamical equilibrium.
In order to fill this gap, we will study the behavior of photon and fermion plasma under VSR Lorentz violating effects in the presence of a heat bath, exploring its properties and in particular study quantum effects at finite temperature.
Hence, the main purpose of the present work is the evaluation of higher-loop corrections to the effective Lagrangian of VSR electrodynamics plasma; these effective functions are important because they generate all thermal Green's functions of the system \cite{Braaten:1989mz,Frenkel:1989br,Taylor:1990ia,Braaten:1991gm}.

In this paper, we examine the behavior of the VSR-modified electrodynamics
in thermodynamical equilibrium within the Matsubara imaginary-time
formalism \cite{kapusta_gale}.
We start Sec.~\ref{sec2} by reviewing the main aspects of the VSR gauge invariance and establishing the dynamics for the fermion and gauge fields in the VSR electrodynamics.
Moreover, we evaluate in detail the three contributions arising from the one-particle irreducible diagrams at one-loop order for the effective Lagrangian.
This analysis will also allow us to address the question of the massive modes stability of VSR photons in the high-temperature regime.
In order to examine the role played by the new gauge couplings generated within the VSR framework at finite temperature, we consider the two-loop contributions to the effective Lagrangian in Sec.~\ref{sec3}.
At this order there is the contribution of two diagrams that are evaluated in the high-temperature regime; this approximation permits us to determine the leading VSR modifications.
In Sec.\ref{conc}, we summarize the results and present our final remarks.


\section{Gauge fields in VSR}
\label{sec2}

The gauge fixed Lagrangian density for the VSR SIM(2) electrodynamics with the respective ghost fields is written as
\begin{align}
{\cal{L}} & =-\frac{1}{4}\tilde{F}_{\mu\nu}\tilde{F}^{\mu\nu}-\frac{1}{2\xi}\left(\tilde{\partial}_{\mu}A^{\mu}\right)^{2}+\tilde{\partial}_{\mu}\overline{c}\tilde{\partial}^{\mu}c+\bar{\psi}\left(i\gamma^{\mu}\nabla_{\mu}\right)\psi \label{eq:3}
\end{align}
where we have chosen the VSR Lorentz condition $\Omega\left[A\right]=\tilde{\partial}_{\mu}A^{\mu}=0$.
It is well known that under this condition the massive character of
the VSR gauge field is manifest by computing its equation of
motion $\left(\square+m^{2}\right)A_{\mu}=J_\mu$.
The most interesting part of this framework is that in the presence of massive modes the photon still has two physical polarization states \cite{Cheon:2009zx}.

The gauge invariant covariant derivative $\nabla_{\mu}$ present in Eq.~\eqref{eq:3} can be determined by making use of the SIM(2) gauge transformation $\delta A_{\mu}=\tilde{\partial}_{\mu}\Lambda$ and imposing the transformation law for a charged field $\delta\left(\nabla_{\mu}\psi\right)=i\Lambda\left(\nabla_{\mu}\psi\right)$. Under these conditions we have that
\begin{equation}
\nabla_{\mu}\psi=D_{\mu}\psi 	+\frac{1}{2} \frac{m^2}{\left(n\cdot D\right)} n_{\mu}\psi, \label{eq:2a}
\end{equation}
where this definition reduces to the wiggle derivative $\tilde{\partial}_{\mu}$ in the noninteracting case. Also we have used the ordinary covariant derivative $D_\mu = \partial_\mu -ie A_\mu $.
Additionally, we can determine the wiggle field strength from
the usual definition $i \left[\nabla_{\mu},\nabla_{\nu}\right]\phi= \tilde{F}_{\mu\nu}\phi$, so that we find
\begin{equation}
\tilde{F}_{\mu\nu}=\tilde{\partial}_{\mu}A_{\nu}-\tilde{\partial}_{\nu}A_{\mu}\label{eq:2}
\end{equation}
 or in terms of the usual derivative
\begin{align}
\tilde{F}_{\mu\nu}=\partial_{\mu}A_{\nu}+\frac{m^{2}}{2}n_{\mu}\left(\frac{1}{\left(n\cdot\partial\right)^{2}}\partial_{\nu}\left(n\cdot A\right)\right)-\mu\leftrightarrow\nu.
\end{align}
Now that we have established the basic features of the VSR electrodynamics needed for a thermal field theory, we shall proceed to compute the partition function for the model.
Moreover, in order to give physical  meaning to the outcomes in perturbation theory we consider the high-temperature limit of the Green's functions; one needs to take into account the hard thermal loops \cite{Braaten:1989mz}.

Some comments about the expression \eqref{eq:2a} are in place.
About the perturbative analysis, the presence of the term $1/\left(n.D\right)$ in \eqref{eq:2a} shows that there are now an infinite number of interactions (in the coupling $e$).
The Feynman rules for these interactions can be obtained within the Wilson lines approach, which express the  respective terms in a convenient form with $N=1,2,3,...$ legs of photon fields \cite{Dunn:2006xk}, making perturbative analysis workable.
Since we are interested in determining the VSR modifications up to the two-loop internal energy density, it suffices to derive the Feynman rules for the $\left\langle \overline{\psi} \psi A \right\rangle $ and $\left\langle \overline{\psi} \psi A A \right\rangle $ vertices.

\subsection{One-loop contribution}

\begin{figure}[t]
\vspace{-1.2cm}
\includegraphics[height=6.5\baselineskip]{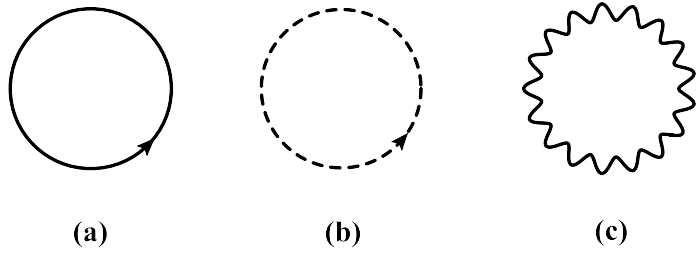}
\centering\caption{Diagrams contributing to the one-loop effective Lagrangian: (a) Fermionic loop, (b) ghost loop, and (c) photon loop.}
\label{oneloop1}
\end{figure}

We start by computing the lowest order contributions to the effective Lagrangian, which will be performed by use of the Matsubara imaginary-time formalism \cite{kapusta_gale}.
At this order, we have three contributions that are presented in Fig.~\ref{oneloop1}.
These diagrams correspond to the fermionic loop (a), ghost
loop (b), and photon loop (c).
The lowest order contributions to the effective Lagrangian are the ring diagrams
\begin{equation}
\mathcal{L}^{(1)}={\mathcal{L}}_{_{(1)}}^{\textrm{photon}}+\mathcal{L}_{_{(1)}}^{\textrm{ghosts}}
+\mathcal{L}_{_{(1)}}^{\textrm{fermions}},\label{eq:4}
\end{equation}
with the following definitions
\begin{align}
\mathcal{L}_{_{(1)}}^{\textrm{photon}} & =-\frac{1}{2}\ln\det\left(M_{\mu\nu}\right),\label{eq:5}\\
\mathcal{L}_{_{(1)}}^{\textrm{ghosts}} & =\ln\det\left(\tilde{\partial}_{\mu}\tilde{\partial}^{\mu}\right),\\
\mathcal{L}_{_{(1)}}^{\textrm{fermions}} & =\ln {\rm det}_{D} \left(i\gamma^{\mu}\tilde{\partial}_{\mu}\right),
\label{eq:6}
\end{align}
where we have introduced the operator $M_{\mu\nu}$ by simplicity
\begin{equation}
M_{\mu\nu}\left(x,y\right)=\left[\eta_{\mu\nu}\tilde{\partial}_{\lambda}\tilde{\partial}^{\lambda}-\Big(1-\frac{1}{\xi}\Big)
\tilde{\partial}_{\mu}\tilde{\partial}_{\nu}\right]\delta^{(4)}\left(x-y\right).
\label{eq:7}
\end{equation}
Notice that in Eq.~\eqref{eq:6} we denoted ${\rm det}_{D}$ as being the determinant over the Dirac matrices and the Hilbert space ($\det$).
Moreover, in the photon contribution we also have a determinant
over the spacetime indices in addition to the Hilbert space ones.
Hence, with some algebra we find the result $\det\left(M_{\mu\nu}\right)=\frac{1}{\xi}\Big(\tilde{\partial}_{\lambda}\tilde{\partial}^{\lambda}\Big)^{4}$, so that
\begin{align}
\mathcal{L}_{_{(1)}}^{\textrm{photon}} & =-\frac{4}{2}\ln\det\left(\tilde{\partial}_{\lambda}\tilde{\partial}^{\lambda}\right),\\
\mathcal{L}_{_{(1)}}^{\textrm{ghosts}} & =\ln\det\left(\tilde{\partial}_{\mu}\tilde{\partial}^{\mu}\right),\\
\mathcal{L}_{_{(1)}}^{\textrm{fermions}} & =\ln {\rm det}_{D}\left(i \gamma^{\mu}\tilde{\partial}_{\mu}\right).
\end{align}

We finally can add the above contribution into Eq.~\eqref{eq:4} and use the imaginary-time formalism to express the respective one-loop contribution as
\begin{align}
\mathcal{L}^{(1)}  & =-\frac{1}{\beta}\underset{n_{_{B}}}{\sumint  }~\ln\left[-\beta^{2}\left(p^{2}-m^{2}\right)\right] +\frac{1}{2\beta}2^{E\left(d/2\right)}
\underset{n_{_{F}}}{\sumint}~ \ln\left(-\beta^{2}\left(p^{2}-m^{2}\right)\right),\label{eq:8}
\end{align}
in which we have used the fact that $\tilde{p}^2 = p^2 +m^2$ and defined the notation for the fermionic and bosonic sum/integral
\begin{equation}
\underset{n_{_{F}}}{\sumint  } \equiv \underset{n_{_{F}}}{\sum  }\frac{d^{\omega -1}p}{\left( 2\pi \right) ^{\omega -1}}, \quad
\underset{n_{_{B}}}{\sumint  } \equiv \underset{n_{_{B}}}{\sum  }\frac{d^{\omega -1}p}{\left( 2\pi \right) ^{\omega -1}}.
\end{equation}
where we should note that the first sum is over $p_{0}=i\omega_{n_{_{F}}}$,
where $\omega_{n_{_{F}}}=\frac{\left(2n+1\right)\pi}{\beta}$ are the
fermionic Matsubara frequencies, and in the second sum $p_{0}=i\omega_{n_{_{B}}}$, where $\omega_{n_{_{B}}}=\frac{2n\pi}{\beta}$ is the bosonic Matsubara frequency.
Moreover, we are employing the irreducible representation for the Dirac
matrices, so that the trace of the identity is given by $tr\left(I\right)=2^{E\left(\omega/2\right)}$, where $E\left(\omega/2\right)$ is the integer part of $\omega/2$.

An important feature of the VSR Lorentz violating effects is present in the
first term of Eq.~\eqref{eq:8}, corresponding to the added contribution from the photon and ghost fields. There, we can observe the presence of a $2/2=1$ factor, signaling the presence of $2$ physical d.o.f., rather than
the $3/2$ factor of a Proca massive gauge field (with $3$
d.o.f.).
This novel description can be ascribed to the modified gauge invariance in the VSR framework, which allows the description of massive modes of a gauge field in terms of a ``massless'' one, hence the number of physical polarization states for these massive modes is 2 as the ordinary massless Maxwell photon.

We start by evaluating first the fermionic sum and
integral from \eqref{eq:8} by means of imaginary-time formalism
\begin{align}
I_{F}   =  \underset{n_{_{F}}}{\sumint  }~\ln\left(-\beta^{2}\left[\left(q_{0}\right)^{2}-\left\vert \mathbf{q} \right\vert ^{2}-m^{2}\right]\right)
  =  2\int\frac{d^{\omega-1}q}{\left(2\pi\right)^{\omega-1}}\ln\left(1+e^{-\beta\omega_{m}}\right),
   \label{eq 0.20}
\end{align}
where $\omega_{m}=\sqrt{\mathbf{q}^{2}+m^{2}}$ and the remaining integral can be solved by using the standard rules of finite temperature integration.
We should emphasize that all temperature-independent parts of \eqref{eq:8}
lead to a divergent result, i.e., the zero-point energy of the vacuum,
which can be subtracted off since it is an unobservable constant.
It is important to remark however that the subtraction of the temperature-independent parts, at any loop order, is related to the process of renormalization \cite{kapusta_gale}.

The renormalizability of nonlocal field theories, in particular with nonlocality in time, is a difficult problem and its general features are well summarized by Marnelius \cite{Marnelius:1974rq}.
Due to the renewed interest in nonlocal field theories, some steps towards the renormalizability of some types of nonlocal models at finite temperature have been considered in the literature \cite{Biswas:2010xq}.
In addition, the issue of renormalizability of VSR invariant theories, at zero temperature, has been addressed in Ref.~\cite{Alfaro:2015fha}, where it is argued that the presence of VSR nonlocal terms $1/\left(n.p\right)$ does not affect  the  renormalizability  of  the model, because they vanish in the large momentum limit, leaving the same ultraviolet behavior as in the Lorentz invariant theories.

For instance, we shall use the well-known result for the fermionic fields
\begin{equation}
\int_{0}^{\infty }\frac{z^{x-1}}{1+e^{z}}dz=\left( 1-2^{1-x}\right) \Gamma
\left( x\right) \zeta \left( x\right) ,
\end{equation}%
as well for the bosonic fields
\begin{equation}
\int_{0}^{\infty }\frac{z^{x-1}}{e^{z}-1}dz=\Gamma \left( x\right) \zeta
\left( x\right) .
\end{equation}%
After some algebra, we can rewrite \eqref{eq 0.20} into the expression
\begin{equation}
I_{F} =\frac{2\beta}{\left(4\pi\right)^{\frac{\omega-1}{2}}}\frac{m^{\omega}}{\Gamma\left(\frac{\omega+1}{2}\right)}
\underset{k=1}{\sum}\left(-1\right)^{k}\int_{1}^{\infty}dw\left(w^{2}-1\right)^{\frac{\omega-1}{2}}e^{-k\beta mw},
\end{equation}
where one may use the representation of the modified Bessel function of the second-kind  \cite{abramowitz_stegun}
\begin{equation}
\frac{\sqrt{\pi }}{\Gamma \left( \frac{\omega +1}{2}\right) }\int_{1}^{\infty }dx\left( x^{2}-1\right) ^{\frac{\omega -1}{2}}e^{-k\beta Mx}
=\Big( \frac{2}{k\beta M}\Big) ^{\frac{\omega }{2}}K_{\frac{\omega }{2}}\left( k\beta M\right) .
\end{equation}%
to solve the above integral, and we finally obtain
\begin{equation}
I_{F}  =\frac{4\beta m^{\omega}}{\left(2\pi\right)^{\frac{\omega}{2}}}\underset{k=1}{\sum}\left(-1\right)^{k}\Big(\frac{1}{k\beta m}\Big)^{\frac{\omega}{2}}K_{\frac{\omega}{2}}\left(k\beta m\right).\label{eq:15}
\end{equation}

Next, the bosonic sum/integral can be computed by the same procedure as the fermionic part; the only change is in performing the Matsubara frequency sum, which reads
\begin{align}
I_{B}&= \underset{n_{_{B}}}{\sumint}~\ln\left(-\beta^{2}\left[ \left(q_{0}\right)^{2}-\left\vert \mathbf{q}\right\vert ^{2} -m^{2}\right]\right) =  2\int\frac{d^{\omega-1}q}{\left(2\pi\right)^{\omega-1}}\ln\left(1-e^{-\beta\omega_{m}}\right)\nonumber \\
 & = -\frac{4\beta m^{\omega}}{\left(2\pi\right)^{\frac{\omega}{2}}}\underset{k=1}{\sum}\Big(\frac{1}{k\beta m}\Big)^{\frac{\omega}{2}}K_{\frac{\omega}{2}}\left(k\beta m\right). \label{eq:10}
\end{align}

Finally, it is easy to verify that the results \eqref{eq:15} and \eqref{eq:10} are well behaved at the limit $\omega\rightarrow4^{+}$. Hence, substituting them into the one-loop expression \eqref{eq:8}, we find
\begin{align}
\mathcal{L}^{(1)}  =\frac{\beta^{-4}}{4\pi^{2}}\underset{k=1}{\sum}\left[1+2\left(-1\right)^{k}\right]\Big(\frac{2m\beta}{k}\Big)^{2}K_{2}\left(k\beta m\right). \label{eq:11}
\end{align}
Although we have obtained a closed-form expression for the one-loop contribution to the effective Lagrangian \eqref{eq:11}, it is worth considering some approximation in order to elucidate the thermal behavior of the system.
In this case, we can explore the thermal properties of Eq.~\eqref{eq:11} assuming the high-temperature case where $\beta m\ll1$, which means that the parameter $m$ should be much less than the thermal energy.
This can be achieved by taking the asymptotic expansion for the modified Bessel function as $\left\vert z\right\vert \rightarrow0$ \cite{abramowitz_stegun}
\begin{equation}
K_{2}\left(z\right)\sim\frac{2}{z^{2}}-\frac{1}{2}+\frac{1}{32}\Big(3-4\gamma-2\ln\frac{z^{2}}{4}\Big)z^{2}+\mathcal{O}(z^{4}).
\end{equation}
where $\gamma$ is the Euler-Mascheroni constant.
Therefore, under this approximation, we can write Eq.\eqref{eq:11} in the leading order in $\beta m$ as
\begin{align}
\mathcal{L}^{(1)}  =\frac{\pi^{2}}{45\beta^{4}} -\frac{7\pi^{2}}{180\beta^{4}}+\frac{3m^{4}}{16\pi^{2}} \ln\Big(\frac{\beta m}{2}\Big)  +\mathcal{O}\left(\beta m\right)^{2}\label{eq:13}
\end{align}
We can see that the first term is due to the photon and ghost field contributions, the second one comes from the fermionic loop, while the third term is due to the leading contribution from VSR effects coming from all fields.
In order to conclude this section, we can compute the internal energy density of the system from the effective Lagrangian
\begin{equation}
u\left(T\right)=-\Big(1+\beta\frac{d}{d\beta}\Big)\mathcal{L}
\end{equation}
Hence, using the result \eqref{eq:13} and disregarding the temperature independent terms, we obtain
\begin{align}
u^{(1)}\left(T\right) \bigg|_{\beta m \ll 1} & =\frac{\pi^{2}}{15}\beta^{-4}-\frac{7\pi^{2}}{60}\beta^{-4}-\frac{3m^{4}}{32\pi^{2}}\ln\big(\beta^{2}m^{2}\big) \label{eq:14}
\end{align}

One can interpret Eq.~\eqref{eq:14} as a generalization of the usual Stefan-Boltzmann law, $u = \sigma T^{4}$. In this case the new constant can be seen as being composite
\begin{equation}
\sigma=\sigma_{\textrm{rad}}+\sigma_{\textrm{ferm}}+\sigma_{\textrm{vsr}}
\end{equation}
with the radiation part $\sigma_{\textrm{rad}}=\frac{\pi^{2}}{15}$ and fermionic part $\sigma_{\textrm{ferm}} =-\frac{7\pi^{2}}{60}$ augmented by an additional contribution
\begin{equation}
\delta\sigma_{\textrm{vsr}} =-\frac{45m^{4}\beta^{4}}{32\pi^{4}}\ln\big(\beta^{2}m^{2}\big)
\end{equation}
which encodes the leading Lorentz violating VSR effects into the Stefan-Boltzmann law.
It is important to stress that the VSR effects give rise to a logarithm correction to the leading $\beta^{-4}$ behavior from the  QED terms in the high-temperature regime.



\section{Two-Loop Effective Lagrangian}
\label{sec3}
\begin{figure}[t]
\vspace{-1.2cm}
\includegraphics[height=6.5\baselineskip]{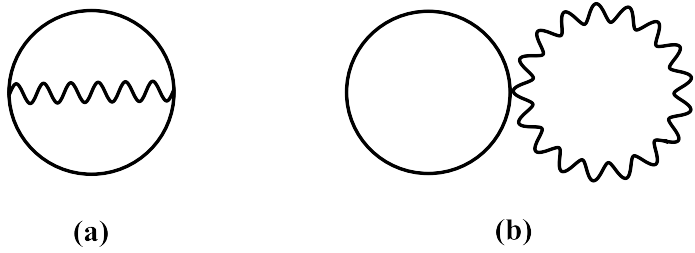}
 \centering\caption{ Diagrams contributing to the two-loop effective Lagrangian: (a) vertex $\left\langle \overline{\psi}\psi A\right\rangle $ and (b) vertex $\left\langle \overline{\psi}\psi A A\right\rangle $.}
\label{twoloop1}
\end{figure}

In order to discuss the role played by the new interactions engendered by VSR Lorentz violation, we shall consider the two-loop diagrams of the SIM(2) electrodynamics. The first contribution is depicted in the panel (a) of Fig.~\ref{twoloop1}.
The 1PI vertex function $\left\langle \overline{\psi} (p_1)\psi (p_2)A(p_3)\right\rangle $ can be obtained from the Lagrangian density \eqref{eq:3} by using a Wilson line approach \cite{Dunn:2006xk}, which results in
\begin{align}
\Lambda^{\mu}\left(p_{1},p_{2},p_{3}\right) & = -ie~\delta^{(4)}(p_{1}+p_{2}+p_{3})\left[\gamma^{\mu}+\frac{m^{2}}{2}
\frac{\left(\gamma.n\right)n^{\mu}}{\left(n.p_{1}\right)\left(n.p_{2}\right)}\right].
\end{align}
Observe that the VSR contribution to this vertex has a nonlocal form.
Moreover, the basic propagators can be computed; for the fermionic field it reads
\begin{equation}
S\left(p\right)=i\frac{\gamma.\tilde{p}}{\tilde{p}^{2} },
\end{equation}
while for the gauge field is written as
\begin{equation}
D^{\mu\nu}\left(p\right)=\frac{-i\eta^{\mu\nu}}{\tilde{k}^{2}},
\end{equation}
in the Feynman gauge $\xi=1$.
Hence, making use of the above Feynman rules, we can write the two-loop contribution from the graph in Fig. 2(a) to the effective Lagrangian,
\begin{align}
\mathcal{L}^{(2)} = & \frac{e^2}{2 \beta^2} \underset{n_{_{F}}}{\sumint  }~~
\underset{m_{_{F}}}{\sumint  }
~~tr\bigg\{\Big[\gamma^{\mu}+\frac{m^{2}}{2}\frac{\left(n.\gamma\right)n^{\mu}}{\left(n.p\right)\left(n.q\right)}\Big]\frac{1}{\gamma.\tilde{p}} \Big[(\gamma^{\nu}+\frac{m^{2}}{2}\frac{\left(n.\gamma\right)n^{\nu}}{\left(n.p\right)\left(n.q\right)}\Big]
\frac{1}{\gamma.\tilde{q}}\bigg\}
\Big(\frac{\eta_{\mu\nu}}{\tilde{k}^{2}}\Big), \label{eq4.2}
\end{align}
where $ k=p-q$ is the photon momentum and both sum/integral are over the
fermionic Matsubara frequencies $p_{0}=i\omega_{n_{_{F}}}=\frac{i\left(2n_{_{F}}+1\right)\pi}{\beta}$
and $q_{0}=i\omega_{m_{_{F}}}=\frac{i\left(2m_{_{F}}+1\right)\pi}{\beta}$.
The computation of finite temperature effects of Eq.~\eqref{eq4.2} follows the same procedure illustrated before.

In comparison to the usual QED, we see the presence of the additional graph  for the two-loop contribution due to the VSR modified gauge invariance, as shown in the panel (b) of Fig.~\ref{twoloop1}.
This diagram is proportional to the 4-point vertex $ \Gamma_{\mu\nu}=\left\langle \overline{\psi}  \psi A _\mu A_\nu \right\rangle $.
However, since the tensor structure of this vertex function has the form  \cite{Dunn:2006xk}
\begin{equation}
\Gamma_{\mu \nu }\left(p_{1},p_{2},p_{3},p_{4}\right)  = -\frac{ie^2 m^2}{2}  \frac{(\gamma.n)n_\mu n_\nu}{(n.p_3)(n.p_4)}
\left[ \frac{1}{(n.p_1)} +\frac{1}{(n.p_2)} -\frac{1}{n.(p_1+p_3)}-\frac{1}{n.(p_1+p_4)}\right],
\end{equation}
it is easy to see that this contribution vanishes identically because the relevant tensor contraction present in the graph is of the type $ \int_{p,q} S \,\Gamma_{\mu \nu } D^{\mu\nu} \sim n_\mu n_\nu \eta^{\mu\nu} =0$.

Computing the trace over the Dirac $\gamma$ matrices with help of the results
$\left\{ \gamma^{\alpha},\gamma^{\beta}\right\} =2\eta^{\alpha \beta}$ and $tr\left(I\right)=2^{E\left(\omega/2\right)}$, we obtain
\begin{align}
\mathcal{L}^{(2)} &=  -\frac{4e^{2}}{\beta^{2}}\underset{n_{_{F}}}{\sumint  }~~
\underset{m_{_{F}}}{\sumint  }~\Big[\frac{\left(\tilde{p}.\tilde{q}\right)-m^{2}}{\tilde{p}^{2}\tilde{q}^{2}\tilde{k}^{2}}
\Big]\nonumber \\
 & =-2e^{2}\left(\Big[I_{F}^{\left(1\right)}\Big]^{2}-2\left[I_{F}^{\left(1\right)}\right]\left[I_{B}^{\left(1\right)}\right]\right)
+\frac{4e^2m^2}{\beta^{2}}
 \underset{n_{_{F}}}{\sumint  }~\underset{m_{_{F}}}{\sumint  } ~\frac{1}{\tilde{p}^{2}\tilde{q}^{2}\tilde{k}^{2}}, \label{eq4.3}
\end{align}
where we have introduced by simplicity of notation the following fermionic and bosonic quantities:
\begin{equation}
I_{F}^{\left(s\right)}  =  \underset{n_{_{F}}}{\sumint  } ~\frac{1}{\left(\tilde{p}^{2}\right)^{s}},\quad
I_{B}^{\left(s\right)}  =  \underset{\ell_{_{B}}}{\sumint  }~\frac{1}{(\tilde{k}^{2})^{s}}.
\end{equation}
In particular, for $s=1$ we can evaluate these expressions by using the finite temperature techniques discussed in the previous section. This analysis yields the result for the fermionic part
\begin{align}
I_{F}^{\left(1\right)}  =\frac{2m^{\omega-2}}{\left(2\pi\right)^{\frac{\omega}{2}}}\underset{k=1}{\sum}\left(-1\right)^{k}\Big(\frac{1}{k\beta m}\Big)^{\frac{\omega-2}{2}}K_{\frac{\omega-2}{2}}\left(k\beta m\right)
\end{align}
and for the bosonic part
\begin{align}
I_{B}^{\left(1\right)}=-\frac{2m^{\omega-2}}{\left(2\pi\right)^{\frac{\omega}{2}}}\underset{n=1}{\sum}\Big(\frac{1}{n\beta m}\Big)^{\frac{\omega-2}{2}}K_{\frac{\omega-2}{2}}\left(n\beta m\right).
\end{align}

Let us now focus on computing the remaining term from Eq.~\eqref{eq4.3}, in particular the frequency sum
\begin{equation}
\mathcal{J }=\frac{1}{\beta^{2}}
 \underset{n_{_{F}}}{\sum  }~\underset{m_{_{F}}}{\sum  } ~\frac{1}{\tilde{p}^{2}\tilde{q}^{2}\tilde{k}^{2}}. \label{eq4.4}
\end{equation}
In order to compute the Matsubara frequency sum present above, which has a complicated structure because of the momentum $k = p-q$, it is helpful to express the Matsubara sum as a contour integral for fermion fields \cite{kapusta_gale}
\begin{equation}
\frac{2}{\beta}\sum_{ n} f\left( p_0 \to i\omega_n\right) = \frac{1}{2\pi i}\oint_C dp_0 ~f\left( p_0\right) \tanh \Big( \frac{ \beta}{2} p_0\Big)
\end{equation}
where $p_0$ is seen as the fourth component of a Minkowski four-vector, the function $  \tanh (p_0)$ has poles at $p_0 =   \frac{i (2n+1)\pi }{\beta}$ and is everywhere else bounded and analytic. Hence, we can rewrite \eqref{eq4.4} in the integral form
\begin{align}
\mathcal{J } & =\frac{1}{4}\oint\frac{dq_{0}}{2\pi i}\frac{1}{\left(q_{0}^{2}-\omega_{q}^{2}\right)}\tanh\Big(\frac{\beta}{2}q_{0}\Big)\oint\frac{dp_{0}}{2\pi i}\left[\frac{\tanh\left(\frac{\beta}{2}p_{0}\right)}{\left(p_{0}^{2}-\omega_{p}^{2}\right)\left(\left(p_{0}-q_{0}\right)^{2}-\omega_{k}^{2}\right)}\right]
\end{align}
where we have defined the massive dispersion relation $\omega_\textbf{q} =\sqrt{\textbf{q}^2 +m^2}$.
As illustration, we can compute first the integral over $p_0$; in this case we see that it has four poles
\begin{equation}
p_{0}=\pm\omega_{p},\quad p_{0}=q_{0}\pm\omega_{k}.
\end{equation}
The evaluation of this contour integral is straightforward, and the integration over $q_0$ of the resulting expression has the following poles:
\begin{equation}
q_{0}=\pm\omega_{q},\quad q_{0}=\pm \omega_{p}\pm\omega_{\textbf{k}}.
\end{equation}
After performing the contour integration for $p_0$ and $q_0$ in the respective poles, we finally arrive at
\begin{align}
\mathcal{J } & =\frac{1}{8\omega_{\textbf{p}}\omega_{\textbf{q}}\omega_{\textbf{k}}}
\Bigg\{\big[n_{_{\textrm{F.D}}}\left(\omega_{p}\right)-n_{_{\textrm{B.E}}}\left(\omega_{\textbf{k}}\right)\big]
\bigg[\frac{4\left(\omega_{\textbf{p}}+\omega_{\textbf{k}}\right)}{\omega_{\textbf{q}}^{2}-\left(\omega_{\textbf{p}}+\omega_{\textbf{k}}\right)^{2}} n_{_{\textrm{F.D}}}\left(\omega_{\textbf{q}}\right) \nonumber \\
& +\frac{1}{\left(\omega_{\textbf{p}}+\omega_{\textbf{k}}\right)^{2}-\omega_{\textbf{q}}^{2}}
\big[n_{_{\textrm{F.D}}}\left(\omega_{\textbf{p}}+\omega_{\textbf{k}}\right)-n_{_{\textrm{F.D}}}
\left(-\omega_{\textbf{p}}-\omega_{\textbf{k}}\right)\big]\bigg]\nonumber \\
 & -\big[n_{_{\textrm{F.D}}}\left(\omega_{\textbf{p}}\right)+n_{_{\textrm{B.E}}}\left(\omega_{\textbf{k}}\right)\big]
 \bigg[\frac{4\left(\omega_{\textbf{p}}-\omega_{\textbf{k}}\right)}{\omega_{\textbf{q}}^{2}-\left(\omega_{\textbf{p}}-\omega_{\textbf{k}}\right)^{2}} n_{_{\textrm{F.D}}}\left(\omega_{\textbf{q}}\right)\nonumber \\
 &+\frac{1}{\left(\omega_{\textbf{p}}-\omega_{\textbf{k}}\right)^{2}-
 \omega_{\textbf{q}}^{2}}\big[n_{_{\textrm{F.D}}}\left(\omega_{\textbf{p}}-\omega_{\textbf{k}}\right)-n_{_{\textrm{F.D}}}
 \left(-\omega_{\textbf{p}}+\omega_{\textbf{k}}\right)\big]\bigg]\Bigg\} \label{eq4.5}
\end{align}
where $n_{_{\textrm{F.D}/\textrm{B.E}}}\left(\omega \right)=\frac{1}{e^{\beta \omega }\pm1}$
is the Fermi-Dirac/Bose-Einstein distribution. We have also used that $\exp\left(\beta q_{0}\right)=\exp\left(i\left(2n_{F}+1\right)\pi\right)=-1$, valid for fermionic fields.
It is important to observe that in Eq.~\eqref{eq4.5} we kept only terms that are quadratic in the occupation numbers, while  those terms that are linear, and terms
that are independent of the occupation numbers are necessarily absorbed by renormalization \cite{kapusta_gale}.
Hence, we are left to compute the momentum integration over spacial components of $p$ and $q$, that is
\begin{equation}
\mathcal{T}= \frac{1}{\beta^{2}}
 \underset{n_{_{F}}}{\sumint  }~\underset{m_{_{F}}}{\sumint  } ~\frac{1}{\tilde{p}^{2}\tilde{q}^{2}\tilde{k}^{2}} = \int\frac{d^{\omega-1}p}{\left(2\pi\right)^{\omega-1}}\frac{d^{\omega-1}q}{\left(2\pi\right)^{\omega-1}} \mathcal{J } \label{eq4.6}
\end{equation}
Although these two-loop integrations are well defined as $\omega\to4^{+}$, the evaluation of the integrals of \eqref{eq4.6} has no analytical solution even with a suitable choice of coordinates.
Nonetheless, we can obtain the leading thermal contribution of the two-loop effective Lagrangian by considering the high-temperature limit $\beta m\ll1$.
In order to highlight the temperature dependence of \eqref{eq4.6}, let us perform the following variable transformation $\left(p,q\right)\to\beta^{-1}\left(p,q\right)$ upon the integrals, which results in
\begin{equation}
\mathcal{T} \left(\beta m\right)=\beta^{6-2\omega}\mathcal{A}\left(\beta m\right)+\beta^{7-2\omega}\mathcal{B}\left(\beta m\right)
\label{eq4.7}
\end{equation}
where
\begin{align}
\mathcal{A}\left(\beta m\right) = &\frac{1}{2}\int\frac{d^{\omega-1}p}{\left(2\pi\right)^{\omega-1}}\frac{d^{\omega-1}q}{\left(2\pi\right)^{\omega-1}}
~\frac{n_{_{\textrm{F.D}}}\left(\omega_{\textbf{q}}\right)}{\omega_{\textbf{p}}\omega_{\textbf{q}}\omega_{\textbf{k}}}\Bigg\{\frac{\left(\omega_{\textbf{p}}+\omega_{\textbf{k}}\right)}{\left(\omega_{\textbf{p}}+\omega_{\textbf{k}}\right)^{2}-\omega_{\textbf{q}}^{2}}\biggl[-n_{_{\textrm{F.D}}}\left(\omega_{\textbf{p}}\right)+n_{_{\textrm{B.E}}}\left(\omega_{\textbf{k}}\right)\biggr]\nonumber \\
 & +
\frac{\left(\omega_{\textbf{p}}-\omega_{\textbf{k}}\right)}
{\left(\omega_{\textbf{p}}-\omega_{\textbf{k}}\right)^{2}-\omega_{\textbf{q}}^{2}}\biggl[n_{_{\textrm{F.D}}}\left(\omega_{\textbf{p}}\right)
+n_{_{\textrm{B.E}}}\left(\omega_{\textbf{k}}\right)\biggr]\Bigg\}
\end{align}
and
\begin{align}
\mathcal{B}\left(\beta m\right) & =\frac{1}{8} \int\frac{d^{\omega-1}p}{\left(2\pi\right)^{\omega-1}}\frac{d^{\omega-1}q}{\left(2\pi\right)^{\omega-1}}
~\frac{1}{\omega_{\textbf{p}}\omega_{\textbf{q}}\omega_{\textbf{k}}}\nonumber\\
&\times\Bigg\{\frac{1}{\left(\omega_{\textbf{p}}+\omega_{\textbf{k}}\right)^{2}-\omega_{q}^{2}}
  \biggl[n_{_{\textrm{F.D}}}\left(\omega_{\textbf{p}}+\omega_{\textbf{k}}\right)
 - n_{_{\textrm{F.D}}}\left(-\omega_{\textbf{p}}-\omega_{\textbf{k}}\right) \biggr]  \left[ n_{_{\textrm{F.D}}}\left(\omega_{\textbf{p}}\right)- n_{_{\textrm{B.E}}}\left(\omega_{\textbf{k}} \right)\right] \nonumber \\
 & -\frac{1}{\left(\omega_{\textbf{p}}-\omega_{\textbf{k}}\right)^{2}-\omega_{\textbf{q}}^{2}}
\biggl[n_{_{\textrm{F.D}}}\left(\omega_{\textbf{p}}-\omega_{\textbf{k}}\right) - n_{_{\textrm{F.D}}}\left(-\omega_{\textbf{p}}+\omega_{\textbf{k}}\right)  \biggr] \left[ n_{_{\textrm{F.D}}}\left(\omega_{\textbf{p}}\right)+ n_{_{\textrm{B.E}}}\left(\omega_{\textbf{k}}\right)\right]\Bigg\}
\end{align}

Besides, we have extracted some of the temperature dependence as in Eq.~\eqref{eq4.7}, the change of the momenta variables has modified the thermal distributions  $n_{_{\textrm{F.D}/\textrm{B.E}}=\frac{1}{e^{\omega_{\textbf{p}}}\pm1}}$, leading to new energies
\begin{equation}
\omega_{\textbf{p}}  =\sqrt{\textbf{p}^{2}+\beta^{2}m^{2}}.
\end{equation}
In this case, it easy to see that the high-temperature limit can be achieved by using the parameter $x=\beta^2 m^2 $ as the coefficient in the Taylor expansion under the condition $\beta m \ll 1$.
Hence, in the high-temperature limit and as $\omega\to4^{+}$ we find that the leading thermal contributions of Eq.~\eqref{eq4.7} are given by
\begin{align}
\mathcal{T}\left(\beta m\right) & \approx\beta^{-2}\mathcal{A}^{\left(0\right)}+\beta^{-1}\mathcal{B}^{\left(0\right)}+\mathcal{O}\left(\beta m\right) \label{eq4.8}
\end{align}
where we have disregarded temperature-independent terms and $\left( \mathcal{A}^{\left(0\right)}, \mathcal{B}^{\left(0\right)}\right)$ are evaluated as the functions $\left( \mathcal{A}, \mathcal{B}\right)$ under the limit $\beta m \to 0$.

Additionally, we should evaluate the remaining contribution of the two-loop effective Lagrangian \eqref{eq4.3} in the high-temperature limit as well.
Thus, under such consideration we have
\begin{align}
 \mathcal{I} &=\left[I_{F}^{\left(1\right)}\right]^{2}-2\left[I_{F}^{\left(1\right)}\right]\left[I_{B}^{\left(1\right)}\right] \nonumber \\
 & =-\frac{\beta^{-4}}{192}-\frac{\beta^{-2}}{48\pi^{2}}m^{2} \ln2
  -\frac{m^{4}}{64\pi^{4}}\biggl[6\ln\pi-3\ln\left(\beta m\right)-6\gamma+3+4\ln2\biggr]\ln\left(\beta m\right)  \label{eq4.9}
\end{align}
where we have used the asymptotic expansion for $\left\vert z\right\vert \rightarrow0$.
\begin{equation}
K_{1}\left(z\right)\sim\frac{1}{z}+\frac{z}{4}\left(2\ln \frac{z}{2}+2\gamma-1 \right)+\mathcal{O}(z^{3}).
\end{equation}

Hence, with the results Eqs.~\eqref{eq4.8} and \eqref{eq4.9}, we can finally write the leading contributions in $m^2$ of the two-loop effective Lagrangian \eqref{eq4.3}
\begin{align}
\mathcal{L}^{(2)}  & = 4e^{2}m^{2}\left(\beta^{-2}\mathcal{A}^{\left(0\right)}+\beta^{-1}\mathcal{B}^{\left(0\right)}\right) +e^{2}\frac{\beta^{-4}}{96}+e^{2}\frac{\beta^{-2}}{24\pi^{2}}m^{2}\ln2\nonumber \\
 & +e^{2}\frac{m^{4}}{32\pi^{4}}\left(6\ln\pi-3\ln\left(\beta m\right)-6\gamma+3+4\ln2\right)\ln\left(\beta m\right)+\mathcal{O}\left(\beta m\right)
\end{align}
from which we can compute the respective internal energy density
\[
u^{\left(2\right)}\left(T\right)=-\left(1+\beta\frac{d}{d\beta}\right)\mathcal{L}^{(2)}
\]
The complete VSR modified energy density is given in terms of the one- and two-loop contributions as
\begin{equation}
u\left(T\right) = u^{\left(1\right)}\left(T\right)+u^{\left(2\right)}\left(T\right)
\end{equation}
Thus, by evaluating the remaining derivatives and discarding the temperature independent terms, we can express the energy density in the high-temperature regime as follows
\begin{align} \label{eq5.0}
u\left(T\right) & =\frac{\pi^{2}}{15}\left[1 -\frac{7 }{4} +\frac{15e^{2}}{32 \pi^{2}}\right]\beta^{-4}+\frac{e^{2}m^{2}}{24\pi^{2}}\left[96\pi^{2}\mathcal{A}^{\left(0\right)}-\ln2\right]\beta^{-2}\nonumber \\
 & -\frac{m^{4}}{32\pi^{2}}\left[3+\frac{2e^{2}}{\pi^{2}}\left(12\ln\pi-3\ln\left[\beta^{2}m^{2}\right]-12\gamma+8\ln2\right)\right]\ln\left[\beta^{2}m^{2}\right]+\mathcal{O}\left(\beta m\right)
\end{align}
We observe that in addition to the VSR logarithmic correction which is already present in the one-loop order, the VSR two-loop contributions give also a  subleading $\beta^{-2}$ correction.
If we associate the VSR parameter to a feasible value for the photon mass $m_\gamma \leq 10^{-18}eV$ \cite{Tanabashi:2018oca}, we see that in the high-temperature regime the two-loop term $e^2 m^2 \beta^{-2}$ is actually greater than the one-loop contribution $m^4 \ln \beta$.
This is a surprising phenomenon since it breaks the hierarchy of the loop expansion.
Nonetheless, these VSR contributions represent a small modification of the leading $\beta^{-4}$ behavior from the known QED terms.

\section{Final remarks}
\label{conc}

In this paper, we presented a study on the thermodynamical properties of the quantum electrodynamics in the VSR Lorentz violating framework  by evaluating systematically the one-loop and two-loop expressions for the effective Lagrangian.
Our aim in this paper was to establish the behavior of the VSR leading effects of some thermal quantities.
We started our analysis by discussing the computation of the effective action at the one-loop order for the case of VSR quantum electrodynamics.
In this approximation, we have only the ring diagram contributions coming from the free gauge, ghosts and fermion fields.
A first comment is based on the number of physical polarization states for the photon: when we add the contributions from the gauge and ghost fields, we see that although the photon propagates massive modes, the number of physical d.o.f. of the photon is preserved in the VSR framework, not presenting a third and longitudinal state as in Proca's massive electrodynamics.
By adding the ring diagram contributions, we obtained the exact expression for the one-loop effective Lagrangian written in terms of a series \eqref{eq:11}.
To elucidate the leading $m^2$ contributions caused by VSR into the thermal content of the system, we have considered the high-temperature regime where $\beta m\ll1$, obtaining thus that the Lorentz violating effects appear as a logarithm correction to the $\beta^{-4}$ behavior for the energy density.

The key mechanism behind the preservation of the number of physical d.o.f. is the VSR modified gauge invariance, in which the gauge field transforms as $\delta A_{\mu}=\tilde{\partial}_{\mu}\Lambda$.
Naturally, this new gauge invariance implies new types of interactions when applied to the QED.
An interesting way to analyze the effects of this VSR coupling between fermions and photons in thermodynamical equilibrium is obtained by evaluating the two-loop effective action.
Although the two-loop contribution from the ordinary coupling can be computed exactly, the expression from the additional contribution due to the VSR gauge coupling is complicated, and has no analytical solution even with a suitable choice of coordinates.
In this case, we have evaluated the two-loop effective Lagrangian by considering the high-temperature limit $\beta m\ll1$, so that we can determine the leading $m^2$ contributions to this thermal quantity.

From the complete expression for the energy density \eqref{eq5.0}, in the high-temperature regime, we see that the VSR contributions yield to interesting modifications from the ordinary QED behavior, corresponding to $m^2 \beta^{-2}$ and $m^4 \ln\beta$ correction terms to the ordinary $\beta^{-4}$ behavior of QED.
Motivated by this result, a subsequent study of the present model would consist of a systematic discussion on nonperturbative phenomena in the VSR electrodynamics at finite temperature.
It is known that infrared divergences are present in higher-order contributions generated by massless modes; such a divergence may be traced back to the dominant high-temperature contribution of the zero mode.
Hence, it would be a nice analysis to study how the VSR effects can modify the infrared divergence profile from QED.

 \subsection*{Acknowledgements}

The authors would like to thank the anonymous referee
for his/her comments and suggestions to improve this
paper.
R.B. acknowledges partial support from Conselho
Nacional de Desenvolvimento Cient\'ifico e Tecnol\'ogico (CNPq Projects No. 304241/2016-4 and No. 421886/2018-8) and Funda\c{c}\~ao de
Amparo \`a Pesquisa do Estado de Minas Gerais (FAPEMIG Project No. APQ-01142-17).


\global\long\def\link#1#2{\href{http://eudml.org/#1}{#2}}
 \global\long\def\doi#1#2{\href{http://dx.doi.org/#1}{#2}}
 \global\long\def\arXiv#1#2{\href{http://arxiv.org/abs/#1}{arXiv:#1 [#2]}}
 \global\long\def\arXivOld#1{\href{http://arxiv.org/abs/#1}{arXiv:#1}}


\end{document}